\documentclass{kapproc} 
\usepackage{latexsym,epsfig,amsfonts}
\kluwerbib

\normallatexbib 
\newcommand{\beq}{\begin{equation}}
\newcommand{\eeq}{\end{equation}}
\newcommand{\be}{\begin{eqnarray}}
\newcommand{\ee}{\end{eqnarray}}
\newcommand{\bdm}{\begin{displaymath}}
\newcommand{\edm}{\end{displaymath}}

\def\la{\mathrel{\mathpalette\fun <}}
\def\ga{\mathrel{\mathpalette\fun >}}
\def\fun#1#2{\lower3.6pt\vbox{\baselineskip0pt\lineskip.9pt
\ialign{$\mathsurround=0pt#1\hfill##\hfil$\crcr#2\crcr\sim\crcr}}}
\def\bce{\begin{center}}
\def\ece{\end{center}}

\newcommand{\AmS}{{\protect\the\textfont2
  A\kern-.1667em\lower.5ex\hbox{M}\kern-.125emS}}

\newskip\humongous \humongous=0pt plus 1000pt minus 1000pt

\newif\ifdtup

\def\deg{\ifmmode ^\circ                
         \else $^\circ$
         \fi
         \hskip -0.1truecm}
\def\degd#1.#2{                         
               \ifmmode {#1^{\hskip 0.05em\circ}\hskip-0.42em.\hskip0.08em#2}
               \else {#1$^{\hskip 0.05em\circ}\hskip-0.42em.\hskip0.08em$#2}
               \fi
              }
\def\mind#1.#2{                         
               \ifmmode {#1^{\hskip 0.05em\prime}\hskip-0.35em.\hskip0.05em#2}
               \else {#1$^{\hskip 0.05em\prime}\hskip-0.35em.\hskip0.05em$#2}
               \fi
              }
\def\secd#1.#2{                         
               \ifmmode {#1^{\prime\prime}\hskip-0.46em.\hskip0.12em#2}
               \else {#1$^{\prime\prime}\hskip-0.46em.\hskip0.12em$#2}
               \fi
              }
\def\timsecd#1.#2{                      
                  \ifmmode {#1^{\rm s}\hskip-0.39em.\hskip0.08em#2}
                  \else {$#1^{\rm s}\hskip-0.39em.\hskip0.08em#2$}
                  \fi
                 }
\def\hms#1h#2m#3s{                      
                  \relax
                  \ifmmode #1^{\rm h}\,#2^{\rm m}\,#3^{\rm s}
                  \else \hbox{$#1^{\rm h}\,#2^{\rm m}\,#3^{\rm s}$}
                  \fi
                 }
\def\dms#1d#2m#3s{                      
                  \relax
                  \ifmmode #1^\circ\,#2^{\prime}\,#3^{\prime\prime}
                  \else \hbox{$#1^\circ\,#2^{\prime}\,#3^{\prime\prime}$}
                  \fi
                 }
\def\dmsd#1d#2m#3.#4s{                  
                      \relax
                      \ifmmode #1^\circ\,#2^{\prime}\,#3^{\prime\prime}
                               \hskip-0.46em.\hskip0.12em#4
                      \else \hbox{$#1^\circ\,#2^{\prime}\,#3^{\prime\prime}
                            \hskip-0.46em.\hskip0.12em#4$}
                      \fi
                     }
\def\hm#1h#2m{                          
              \relax
              \ifmmode #1^{rm h}\,#2^{\rm m}
              \else \hbox{$#1^{\rm h}\,#2^{\rm m}$}
              \fi
             }
\def\dm#1d#2m{                          
              \relax
              \ifmmode #1^\circ\,#2^{\prime}
              \else \hbox{$#1^\circ\,#2^{\prime}$}
              \fi
             }
\def\hmsd#1h#2m#3.#4s{                  
                      \relax
                      \ifmmode #1^{\rm h}\,#2^{\rm m}\,#3^{\rm s}
                               \hskip-0.39em.\hskip0.08em#4
                      \else \hbox{$#1^{\rm h}\,#2^{\rm m}\,#3^{\rm s}
                            \hskip-0.39em.\hskip0.08em#4$}
                      \fi
                     }
\def\hmd#1h#2.#3m{                  
                  \relax
                  \ifmmode #1^{\rm h}\,#2^{\rm m}
                           \hskip-0.55em.\hskip0.22em#3
                  \else \hbox{$#1^{\rm h}\,#2^{\rm m}
                        \hskip-0.55em.\hskip0.22em#3$}
                  \fi
                 }
\def\mg{\relax                          
        \ifmmode ^{\rm m}
        \else $^{\rm m}$
        \fi
       }
\def\mgd#1.#2{                          
              \relax
              \ifmmode #1^{\rm m}
                       \hskip-0.55em.\hskip0.22em#2
              \else \hbox{#1$^{\rm m}
                    \hskip-0.55em.\hskip0.22em$#2}
              \fi
             }

\def\la{\mathrel{\hbox{\rlap{\hbox{\lower4pt\hbox{$\sim$}}}\hbox{$<$}}}}
\def\ga{\mathrel{\hbox{\rlap{\hbox{\lower4pt\hbox{$\sim$}}}\hbox{$>$}}}}

\def\unitspace{\;}                      

\def\un#1{\ifmmode \unitspace\mbox{\rm #1} 
          \else $\unitspace$#1
          \fi}
\def\pun#1#2{\ifmmode \unitspace\mbox{\rm #1}^{#2} 
             \else $\unitspace$#1$^{#2}$
             \fi}

\def\Lsun{\ifmmode \un{L}_{\odot}     
          \else $\un{L}_{\odot}$
          \fi}
\def\Msun{\ifmmode \un{M}_{\odot}     
          \else $\un{M}_{\odot}$
          \fi}
\def\mum{\ifmmode \unitspace\mu\mbox{\rm m} 
         \else $\unitspace\mu$m
         \fi}
\def\sqarcsec{\ifmmode \unitspace\Box''    
              \else $\unitspace\Box''$     
              \fi} 

\def\Bp{\relax                            
        \ifmmode B_{||}                   
        \else $B_{||}$
        \fi}
\def\Bt{\relax                            
        \ifmmode B\!_{\perp}              
        \else $B\!_{\perp}$               
        \fi}
\def\Gcr{\relax                           
         \ifmmode \Gamma\!_{\rm cr}       
         \else $\Gamma\!_{\rm cr}$
         \fi}
\def\ICII{\relax                          
          \ifmmode I_{[\CII]}             
          \else $I_{[\CII]}$
          \fi}
\def\LHtwo{\relax                                 
           \ifmmode L_{\mbox{\rm\scriptsize H}_2} 
           \else $L_{\mbox{\rm\scriptsize H}_2}$  
           \fi}
\def\LLya{\relax                          
          \ifmmode L_{{\rm Ly}\,\alpha}   
          \else $L_{{\rm Ly}\,\alpha}$
          \fi}
\def\MHtwo{\relax                                 
           \ifmmode M_{\mbox{\rm\scriptsize H}_2} 
           \else $M_{\mbox{\rm\scriptsize H}_2}$  
           \fi}
\def\MHtwodot{\relax                                       
              \ifmmode \dot{M}_{\mbox{\rm\scriptsize H}_2} 
              \else $\dot{M}_{\mbox{\rm\scriptsize H}_2}$  
              \fi}                                         
\def\Mstardot{\relax                      
              \ifmmode \dot{M}_{\ast}     
              \else $\dot{M}_{\ast}$      
              \fi}
\def\nHI{\relax                                      
         \ifmmode n_{\mbox{\scriptsize\rm H\,\sc I}} 
         \else $n_{\mbox{\scriptsize\rm H\,\sc I}}$
         \fi}
\def\nHtwo{\relax                                
           \ifmmode n_{{\mbox{\scriptsize H}}_2} 
           \else $n_{{\mbox{\scriptsize H}}_2}$  
           \fi}
\def\rhostardot{\relax                         
                \ifmmode \dot{\rho}_{\ast}     
                \else $\dot{\rho}_{\ast}$      
                \fi}
\def\rhoZdot{\relax                          
             \ifmmode \dot{\rho}_{\rm Z}     
             \else $\dot{\rho}_{\rm Z}$      
             \fi}

\def\sou#1#2{\relax                       
             \ifmmode {\rm #1}\,{\rm #2}  
             \else #1$\,$#2
             \fi}

\def\qu#1#2{\relax                          
            \ifmmode #1_{\rm #2}            
            \else $#1_{\rm #2}$
            \fi}
\def\deg{\ifmmode ^\circ                
         \else $^\circ$
         \fi
         \hskip -0.1truecm}
\def\degd#1.#2{                         
               \ifmmode {#1^{\hskip 0.05em\circ}\hskip-0.42em.\hskip0.08em#2}
               \else {#1$^{\hskip 0.05em\circ}\hskip-0.42em.\hskip0.08em$#2}
               \fi
              }
\def\mind#1.#2{                         
               \ifmmode {#1^{\hskip 0.05em\prime}\hskip-0.35em.\hskip0.05em#2}
               \else {#1$^{\hskip 0.05em\prime}\hskip-0.35em.\hskip0.05em$#2}
               \fi
              }
\def\secd#1.#2{                         
               \ifmmode {#1^{\prime\prime}\hskip-0.46em.\hskip0.12em#2}
               \else {#1$^{\prime\prime}\hskip-0.46em.\hskip0.12em$#2}
               \fi
              }
\def\timsecd#1.#2{                      
                  \ifmmode {#1^{\rm s}\hskip-0.39em.\hskip0.08em#2}
                  \else {$#1^{\rm s}\hskip-0.39em.\hskip0.08em#2$}
                  \fi
                 }
\def\hms#1h#2m#3s{                      
                  \relax
                  \ifmmode #1^{\rm h}\,#2^{\rm m}\,#3^{\rm s}
                  \else \hbox{$#1^{\rm h}\,#2^{\rm m}\,#3^{\rm s}$}
                  \fi
                 }
\def\dms#1d#2m#3s{                      
                  \relax
                  \ifmmode #1^\circ\,#2^{\prime}\,#3^{\prime\prime}
                  \else \hbox{$#1^\circ\,#2^{\prime}\,#3^{\prime\prime}$}
                  \fi
                 }
\def\dmsd#1d#2m#3.#4s{                  
                      \relax
                      \ifmmode #1^\circ\,#2^{\prime}\,#3^{\prime\prime}
                               \hskip-0.46em.\hskip0.12em#4
                      \else \hbox{$#1^\circ\,#2^{\prime}\,#3^{\prime\prime}
                            \hskip-0.46em.\hskip0.12em#4$}
                      \fi
                     }
\def\hm#1h#2m{                          
              \relax
              \ifmmode #1^{rm h}\,#2^{\rm m}
              \else \hbox{$#1^{\rm h}\,#2^{\rm m}$}
              \fi
             }
\def\dm#1d#2m{                          
              \relax
              \ifmmode #1^\circ\,#2^{\prime}
              \else \hbox{$#1^\circ\,#2^{\prime}$}
              \fi
             }
\def\hmsd#1h#2m#3.#4s{                  
                      \relax
                      \ifmmode #1^{\rm h}\,#2^{\rm m}\,#3^{\rm s}
                               \hskip-0.39em.\hskip0.08em#4
                      \else \hbox{$#1^{\rm h}\,#2^{\rm m}\,#3^{\rm s}
                            \hskip-0.39em.\hskip0.08em#4$}
                      \fi
                     }
\def\hmd#1h#2.#3m{                  
                  \relax
                  \ifmmode #1^{\rm h}\,#2^{\rm m}
                           \hskip-0.55em.\hskip0.22em#3
                  \else \hbox{$#1^{\rm h}\,#2^{\rm m}
                        \hskip-0.55em.\hskip0.22em#3$}
                  \fi
                 }
\def\mg{\relax                          
        \ifmmode ^{\rm m}
        \else $^{\rm m}$
        \fi
       }
\def\mgd#1.#2{                          
              \relax
              \ifmmode #1^{\rm m}
                       \hskip-0.55em.\hskip0.22em#2
              \else \hbox{#1$^{\rm m}
                    \hskip-0.55em.\hskip0.22em$#2}
              \fi
             }

\def\la{\mathrel{\hbox{\rlap{\hbox{\lower4pt\hbox{$\sim$}}}\hbox{$<$}}}}
\def\ga{\mathrel{\hbox{\rlap{\hbox{\lower4pt\hbox{$\sim$}}}\hbox{$>$}}}}

\def\unitspace{\;}                      

\def\un#1{\ifmmode \unitspace\mbox{\rm #1} 
          \else $\unitspace$#1
          \fi}
\def\pun#1#2{\ifmmode \unitspace\mbox{\rm #1}^{#2} 
             \else $\unitspace$#1$^{#2}$
             \fi}

\def\Lsun{\ifmmode \un{L}_{\odot}     
          \else $\un{L}_{\odot}$
          \fi}
\def\Msun{\ifmmode \un{M}_{\odot}     
          \else $\un{M}_{\odot}$
          \fi}
\def\mum{\ifmmode \unitspace\mu\mbox{\rm m} 
         \else $\unitspace\mu$m
         \fi}
\def\sqarcsec{\ifmmode \unitspace\Box''    
              \else $\unitspace\Box''$     
              \fi} 

\def\Bp{\relax                            
        \ifmmode B_{||}                   
        \else $B_{||}$
        \fi}
\def\Bt{\relax                            
        \ifmmode B\!_{\perp}              
        \else $B\!_{\perp}$               
        \fi}
\def\Gcr{\relax                           
         \ifmmode \Gamma\!_{\rm cr}       
         \else $\Gamma\!_{\rm cr}$
         \fi}
\def\ICII{\relax                          
          \ifmmode I_{[\CII]}             
          \else $I_{[\CII]}$
          \fi}
\def\LHtwo{\relax                                 
           \ifmmode L_{\mbox{\rm\scriptsize H}_2} 
           \else $L_{\mbox{\rm\scriptsize H}_2}$  
           \fi}
\def\LLya{\relax                          
          \ifmmode L_{{\rm Ly}\,\alpha}   
          \else $L_{{\rm Ly}\,\alpha}$
          \fi}
\def\MHtwo{\relax                                 
           \ifmmode M_{\mbox{\rm\scriptsize H}_2} 
           \else $M_{\mbox{\rm\scriptsize H}_2}$  
           \fi}
\def\MHtwodot{\relax                                       
              \ifmmode \dot{M}_{\mbox{\rm\scriptsize H}_2} 
              \else $\dot{M}_{\mbox{\rm\scriptsize H}_2}$  
              \fi}                                         
\def\Mstardot{\relax                      
              \ifmmode \dot{M}_{\ast}     
              \else $\dot{M}_{\ast}$      
              \fi}
\def\nHI{\relax                                      
         \ifmmode n_{\mbox{\scriptsize\rm H\,\sc I}} 
         \else $n_{\mbox{\scriptsize\rm H\,\sc I}}$
         \fi}
\def\nHtwo{\relax                                
           \ifmmode n_{{\mbox{\scriptsize H}}_2} 
           \else $n_{{\mbox{\scriptsize H}}_2}$  
           \fi}
\def\rhostardot{\relax                         
                \ifmmode \dot{\rho}_{\ast}     
                \else $\dot{\rho}_{\ast}$      
                \fi}
\def\rhoZdot{\relax                          
             \ifmmode \dot{\rho}_{\rm Z}     
             \else $\dot{\rho}_{\rm Z}$      
             \fi}

\def\sou#1#2{\relax                       
             \ifmmode {\rm #1}\,{\rm #2}  
             \else #1$\,$#2
             \fi}

\def\qu#1#2{\relax                          
            \ifmmode #1_{\rm #2}            
            \else $#1_{\rm #2}$
            \fi}

\def\CO#1{\ifnum#1=0                    
           \ifmmode \mbox{\rm CO}
           \else {\rm CO}
           \fi
          \else
           \ifnum#1<15
            \ifmmode ^{#1}\mbox{\rm CO}
            \else $^{#1}${\rm CO}
            \fi
           \else
            \ifmmode \mbox{\rm C}^{#1}\mbox{\rm O}
            \else {\rm C}$^{#1}${\rm O}
            \fi
           \fi
          \fi}

\def\COp{\ifmmode \mbox{\rm CO}^+           
         \else {\rm CO}$^+$                 
         \fi}

\def\CS#1{\ifnum#1=0                    
           \ifmmode \mbox{\rm CS}
           \else {\rm CS}
           \fi
          \else
           \ifnum#1<15
            \ifmmode ^{#1}\mbox{\rm CS}
            \else $^{#1}${\rm CS}
            \fi
           \else
            \ifmmode \mbox{\rm C}^{#1}\mbox{\rm S}
            \else {\rm C}$^{#1}${\rm S}
            \fi
           \fi
          \fi}

\def\HCOp{\ifmmode \mbox{\rm HCO}^+          
          \else {\rm HCO}$^+$                
          \fi}
\def\Hthreep{\ifmmode \mbox{\rm H}_3^+         
             \else {\rm H}$_3^+$               
             \fi}

\def\Htwo{\ifmmode \mbox{\rm H}_2              
          \else {\rm H}$_2$                    
          \fi}

\def\HtwoO{\ifmmode \mbox{\rm H}_2\mbox{\rm O} 
           \else {\rm H}$_2${\rm O}            
           \fi}

\def\ion#1#2{\ifmmode \mbox{{\rm #1}}\,\mbox{{\sc #2}} 
        \else {\rm #1}$\,${\sc #2}
        \fi}

\def\rec#1#2{\if#2a                            
              \ifmmode \mbox{{\rm #1}}\alpha   
              \else {\rm #1}$\alpha$
              \fi
             \fi
             \if#2b
              \ifmmode \mbox{{\rm #1}}\beta
              \else {\rm #1}$\beta$
              \fi
             \fi
             \if#2g
              \ifmmode \mbox{{\rm #1}}\gamma
              \else {\rm #1}$\gamma$
              \fi
             \fi}

\newcommand{\eqref}[1]{Eq.~$\left(\protect\ref{#1}\right)$}

\begin{document}

\articletitle{Large-Scale Environmental Effects of the Cluster Distribution}

\author{Manolis Plionis}
\affil{Institute of Astronomy \& Astrophysics, National Observatory of
Athens, Lofos Koufou, Palaia Penteli, 152 36 Athens, Greece}
\email{plionis@sapfo.astro.noa.gr}

\begin{abstract}
Using the APM cluster distribution we find interesting alignment effects:
{\em (1)} Cluster substructure is strongly correlated with the
tendency of clusters to be aligned with their nearest neighbour and in
general with the nearby clusters that belong to the same
supercluster, {\em (2)} Clusters belonging in superclusters show a
statistical significant tendency to be aligned with the major axis
orientation of their parent supercluster.
Furthermore we find that dynamically young clusters are more 
clustered than the overall cluster population.
These are strong indications that cluster develop in a hierarchical fashion by
merging along the large-scale filamentary superclusters within
which they are embedded.
\end{abstract}

\begin{keywords}
galaxies: clusters - cosmology: general - large-scale structure of 
universe
\end{keywords}

\section{Introduction}
Galaxy clusters are very important for Cosmological studies because:
(a) being the largest bound structures in the universe and
containing hundreds of galaxies and hot X-ray emitting gas, they
can be detected at large distances; (b) they are closed systems with
a mixture of matter which is representative of the whole Universe. 
Therefore, they
appear to be ideal tools for studying the relative abundance of
different types of matter and for testing theories of structure formation 
(cf. B${\rm \ddot{o}}$hringer \cite{bori},
Schindler \cite{Sch00}, Borgani \& Guzzo \cite{borg}).

Below I discuss a few issues related to cluster dynamics and the 
cluster large-scale environment and by using the APM cluster catalogue 
(Dalton {\em et al} \cite{Dal97}) I will present evidence that the are
closely related.

\subsection{Cluster Internal Dynamics \& Cosmology}
One of the interesting properties of galaxy clusters is the relation
between their dynamical state and the underlying cosmology. 
Although the physics of cluster formation is complicated 
(cf. Sarazin \cite{Sar01}), it is expected that in an open or a flat with
vacuum-energy contribution universe, clustering effectively freezes 
at high redshifts (for example in an open model, $z\simeq\;\Omega_{m}^{-1}-1$) 
and thus clusters today should appear more 
relaxed with weak or no indications of substructure. 
Instead, in a critical density model, such systems continue to form even today
and should appear to be dynamically active (cf. Richstone, Loeb \&
Turner \cite{Ri}, Evrard {\em et al.} \cite{Ev}, Lacey \& Cole \cite{Lac}).
Therefore by studying the relative evolution of cluster physical
properties, between distant and nearby clusters, one can attempt to 
extract cosmological information.

Such a task is however hampered by, at least, two facts:
\begin{itemize}
\item {\it Ambiguity in identifying cluster substructure:}
One has to deal with the issue of unambiguously identifying
cluster substructure, since projection effects in the optical
can conspire to make
cluster images appear having multiple peaks/substructure.
Alot of work has been devoted in attempts to find criteria and methods
to identify cluster substructure (see references in
Kolokotronis {\em et al} \cite{Kol}). 
It is evident from all the available studies
that there is neither agreement on the methods 
utilised nor on the exact frequency of clusters having substructure. 
\item{\it Unknown physics of cluster merging:}
The clear-cut theoretical expectations regarding the fraction of
clusters expected to be relaxed in different cosmological backgrounds
break-down due to the complicated
physics of cluster merging (cf. Sarazin \cite{Sar01})
and especially due to the uncertainty of the
post-merging relaxation time. In other words, identifying a cluster
with significant substructure does not necessarily mean that this
cluster is dynamically young, but could reflect a relatively 
ancient merging that has not relaxed yet to an equilibrium configuration. 
\end{itemize}

It has been realized that the optimum approach in detecting clusters
and studying their dynamical state is by using multiwavlength data.
For example, the cluster X--ray emission is proportional 
to the square of the gas density (rather than just density in the optical) 
and therefore it is centrally concentrated, a fact which minimises 
projection effects (cf. Sarazin \cite{Sar88}, Schindler \cite{Sch99}).
The advantage of using optical data is the shear size of the available cluster 
catalogues and thus the statistical significance of the emanating
results. 
However, as we discussed previously, it seems that in order to take 
advantage of the different rates of cluster evolution in the different 
cosmological backgrounds one needs {\em (a)} to find criteria of recent
cluster merging and 
{\em (b)} calibrate the results using high-resolution cosmological
hydro simulation, which will provide the expectations of the different
cosmological models.

Such criteria have been born out of numerical simulations
(cf. Roettiger {\em et al.} \cite{Roe93} \cite{Roe99}) 
and are based on the use of 
multiwavlength data, especially optical and X-ray data but radio as
well (cf. Zabuldoff \& Zaritsky \cite{Zab}, Schindler \cite{Sch99}).
The criteria are based
on the fact that gas is collisional while galaxies are not and
therefore during the merger of two clumps, containing galaxies and gas,
we expect: ({\it 1}) a difference in the spatial positions of the 
highest peak in the galaxy and gas distribution,
({\it 2}) the X-ray emitting gas, due to
compression along the merging direction, to be elongated
perpendicularly along this direction and
({\it 3}) temperature gradients to develop due to
the compression and subsequent shock heating of the gas.

\subsection{Cluster Alignments \& Formation Processes}
Another interesting observable, that was thought initially to provide 
strong constraints on theories of galaxy formation, 
is the tendency of clusters to be aligned with their nearest 
neighbour as well as with other clusters that reside in the same 
supercluster (cf. Binggeli \cite{Bin}, West \cite{We89}, 
Plionis \cite{Plio94}). Analytical work of Bond \cite{Bond86} \cite{Bond87}
in which clusters were identified as peaks of an initial Gaussian random 
field, has shown that such alignments, expected naturally to occur in 
"top-down" scenarios, are also found in hierarchical clustering models of 
structure formation like the CDM. These results were 
corroborated with the use of high-resolution N-body simulations by West
{\em et al} \cite{We91}, Splinter {\em et al} \cite{Spli} 
and Onuora \& Thomas \cite{Onu}.
This fact has been explained as the result of an interesting 
property of Gaussian random fields that occurs for 
a wide range of initial conditions and which is the "cross-talk" between 
density fluctuations on different scales. This property is apparently 
also the cause of the observed filamentariness
observed not only in "pancake" models but also in hierarchical models 
of structure formation;
the strength of the effect, however, differs from model to model. 

There is strong evidence that the brightest galaxy (BCGs) in
clusters are aligned
with the orientation of their parent cluster and even with the
orientation of the large-scale filamentary structure within which they
are embedded (cf. Struble \cite{Str90},
West \cite{We94}, Fuller, West \& Bridges \cite{Fu}). 
Furthermore, there is conflicting evidence regarding the
alignment of cluster galaxies in general with the orientation of their
parent cluster (cf. Djorgovski \cite{Dj}, van Kampen \& Rhee \cite{Kamp}, 
Trevese, Cirimele \& Flin \cite{Tre}). It may be that general galaxy
alignments may be present in forming, dynamically young, clusters, while
in relaxed clusters violent and other relaxation
processes may erase such alignment features.
Such seems to be the case of the Abell 85/87/89 complex 
(see Durret {\em et al} \cite{Durr}) and of Abell 521, a
cluster at $z\simeq 0.25$ which is forming at the intersection of two filaments
(Arnaud {\em et al} \cite{Arn}). 

Using wide-field CFHT imaging data of A521, Plionis, Maurogordato \&
Benoist ({\em in preparation}) have
found statistical significant alignments not only of the predominantly bright 
but also of fainter
galaxies with the major axis direction of the cluster (figure 1).
\begin{figure}[t]
\begin{center}
\psfig{figure=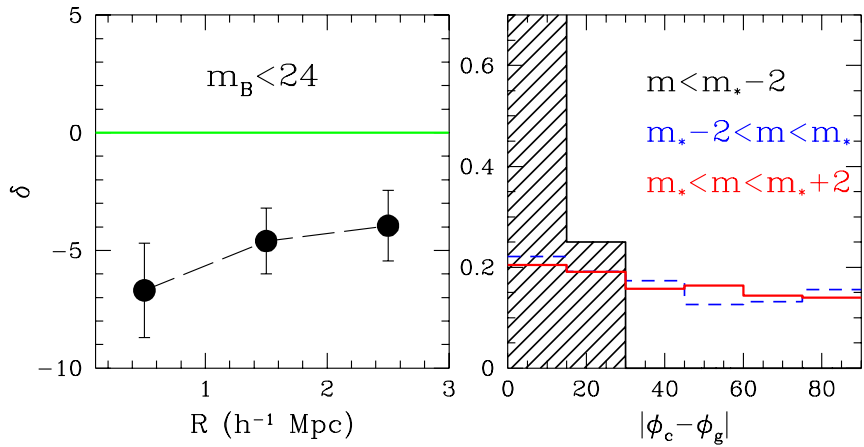,height=2.5in}
\end{center}
\caption{Left: The alignment signal of all
galaxies within 3 $h^{-1}$ Mpc of the cD galaxy of A521 as a function of
distance from it. Right: Frequency distribution of the
misalignment angle between member galaxy and A521 orientations for 3
magnitude bins.}
\end{figure}
It is interesting that the Position angle of the cluster coincides with 
the direction to the nearest Abell cluster (see figure 2).
\begin{figure}[t]
\psfig{figure=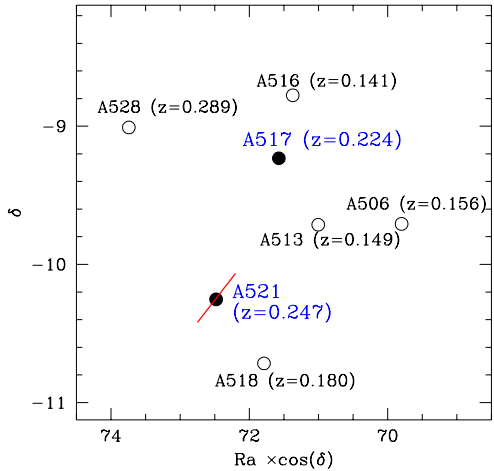,height=2.6in} 
\hfill \parbox[b]{4.5cm}{\caption{The large scale environment
surrounding Abell 521. The major axis direction of A521 is pointing
towards its nearest neighbour A517.}}
\end{figure}
Within the framework of hierarchical clustering, the
anisotropic merger scenario of West \cite{We94}, in which clusters
form by accreting material along the filamentary structure within
which they are embedded, provides an interesting explanation of such
alignments as well as of the observed strong alignment of BCGs 
with their parent cluster orientation.
Evidence supporting this scenario was presented in West, Jones \& 
Forman \cite{We95}
in which they found, using Einstein data, that cluster substructures
seem to be aligned with the orientation of their parent cluster and with the
nearest-neighbouring cluster (see also Novikov {\em et al} \cite{Nov}). 
Such effect has
been observed also in numerical simulations of cluster formation for a
variety of power-spectra (van Haarlem \& van de Weygaert \cite{Haa})

\section{Methodology}
The APM cluster catalogue is based on the APM galaxy survey which
 covers an area of 4300 square degrees in the southern
sky containing about 2.5 million galaxies brighter
than a magnitude limit of $b_{J}=20.5$ (for details see Maddox {\em et
 al.} \cite{Madd}).
Dalton {\em et al.} \cite{Dal97} 
applied an object cluster finding algorithm to the
APM galaxy data using a search radius of  $0.75 \; h^{-1}$ Mpc in order
to minimize projection effects, and so produced a list 
of 957 clusters with $z_{est} < 0.13$. 
Out of these 309 ($\sim 32\%$) are ACO clusters, while 374 
($\sim 39\%$) have measured redshifts (179 of these are ACO clusters).
The APM clusters that are not in the ACO list are relatively poorer
systems than the Abell clusters, as we have verified comparing their 
APM richness's.

Below, I briefly present the methods used to determine the dynamical
state of clusters, their shape, orientation and alignment.

\subsection{Substructure Measures}
As our indicator of cluster substructure of the optical APM data we use
the shift of the center-of-mass 
position as a function of density threshold above which it is
estimated, $sc$ (cf. Evrard {\em et al.} \cite{Ev} and Mohr {\em et al.}
\cite{Mohr}). Kolokotronis {\em et al}  \cite{Kol}, used APM
and X-ray (ROSAT pointed observations) data for 22 clusters and
calibrated this method. Only in $\sim 20\%$ of the
clusters that they studied did they find projection effects in
the optical that altered the X-ray definition of substructure. 
They concluded that a large and significant value of center-of-mass shift
is a clear indication of substructure in the optical APM data (see also
Plionis \cite{Plio01}). 

The significance of the centroid variations to the presence of 
background contamination and random density fluctuations are quantified using 
Monte Carlo cluster simulations in which, by construction, there is no
substructure. For each APM cluster a series of simulated
clusters is produced having the same shape parameters, same 
number of observed galaxies as well as 
a random distribution of background galaxies, determined by the
distance of the cluster and the APM selection function. 
The simulated galaxy distribution follows the usual King-like profile,
which characterizes equilibrium configurations.
Naturally, we expect the simulated clusters to generate
small $sc$'s and in any case insignificant shifts.
Therefore, for each optical cluster, 1000 such Monte-Carlo clusters are
generated and we derive $\langle sc \rangle_{\rm sim}$
as a function of the same density thresholds
as in the real cluster case. Then, we calculate the quantity:
\begin{equation}\label{eq:sig}
\sigma =\frac{\langle sc \rangle_{\rm o} - \langle sc 
\rangle_{\rm sim}}{\sigma_{\rm sim}}\;,
\end{equation}
where $\langle sc \rangle_{\rm o}$
is the centroid shift of the real APM cluster. $\sigma$ 
is a measure of the significance of real centroid shifts
as compared to the simulated, substructure-free clusters, having the
same structural and density parameters as the real cluster.

A further possible substructure identification procedure is based
on a friend-of-friends 
algorithm, applied on 3 overdensity thresholds of each cluster
(for details see Kolokotronis {\em et al.} \cite{Kol}. Three
categories are identified, based on the subgroup multiplicity and
size: {\em (a)} No substructure (unimodal), {\em (b)} 
Weak substructure (multipole groups but with total group mass 
$\le 25\%$ of main), 
{\em (c)} Strong substructure (multipole groups but with mass $>
25\%$ of main). 

\subsection{Shape \& Alignment Measures}

To estimate the cluster parameters we use the familiar moments of 
inertia method, with 
$I_{11}=\sum\ w_{i}(r_{i}^{2}-x_{i}^{2})$,
$I_{22}=\sum\ w_{i}(r_{i}^{2}-y_{i}^{2})$,
$I_{12}=I_{21}=-\sum\ w_{i}x_{i}y_{i}$,
where $x_i$ 
and $y_i$ are the Cartesian coordinates of the galaxies and 
$w_i$ is their weight. We, then diagonalize 
the inertia tensor solving the basic equation:
\begin{equation}\label{eq:diag}
det(I_{ij}-\lambda^{2} \; M_{2})=0 ,
\end{equation}
where $M_{2}$ is the $2 \times 2$ unit matrix. The cluster ellipticity
is given by
$\epsilon=1-\frac{\lambda_2}{\lambda_1}$, where $\lambda_i$ are the
 positive eigenvalues with $(\lambda_1>\lambda_2)$.

This method can be applied to the data using either the discrete or smoothed
distribution of galaxies
(for details see Basilakos, Plionis \& Maddox \cite{bas00}).

In order to test whether there is any significant bias and 
tendency of the position 
angles to cluster around particular values we estimate their Fourier
transform: $C_{n} = \sqrt(2/N) \sum \cos 2n\theta$, 
$S_{n} = \sqrt(2/N) \sum \sin 2n\theta$.
If the cluster position angles, $\theta$, are uniformly distributed 
between $0^{\circ}$ 
and $180^{\circ}$, then both $C_{n}$ and $S_{n}$ have zero mean and unit 
standard deviation. Therefore large values ($>2.5$) indicate significant 
deviation from isotropy.

In order to investigate the alignment between cluster orientations,
we define the relative position angle
between cluster pairs by, $\delta \phi_{i,j}\equiv |\theta_i - \theta_j|$. 
In an isotropic distribution we will have
$\langle \delta \phi_{i,j} \rangle \simeq 45^{\circ}$. 
A significant deviation from this would be an indication of an 
anisotropic distribution which can be quantified by 
(Struble \& Peebles \cite{Str95}):
\begin{equation}\label{eq:alin}
\delta=\sum_{i=1}^{N}\frac{\delta\phi_{i,j}}{N}-45
\end{equation}
In an isotropic distribution we have $\langle \delta \rangle \simeq 0$, while
the standard deviation is given by $\sigma=90/\sqrt{12 N}$. 
A significantly negative value of $\delta$ would indicate alignment and 
a positive misalignment.

\section{APM Cluster Substructure \& Alignments}
Applying the above methodology to the $\sim 900$ APM clusters we find that
about 30\% of clusters have significant ($> 3 \sigma$)
substructure. 
Note that defining 
as having significant substructure those clusters with $\sigma
>2$ or 2.5 increases the fraction to $\sim$ 50\% and 40\%
respectively. 
Alternatively, if we apply the {\em subgroup} categorization procedure
we find that $\sim$53\% of the APM clusters show strong indications of
substructure, which would point that, for consistency among the two
methods, a limit of $\sigma \simeq 2.5$ would be required above which
substructure should be considered significant.

A necessary prerequisite for alignment analyses is that there is no
orientation bias in the distribution of estimated position angles.
In the lower panel of figure 3 we present the corresponding
distribution for the APM clusters. 
It is evident that the distribution is isotropic, as
it is also quantified by the Fourier analysis. In the upper-panel of
figure 3 we present the distribution of relative position angles, 
$\delta\phi$, between APM
nearest-neighbours for two separation limits (one for all separations
and one for separations $< 10 \; h^{-1}$ Mpc). It is
evident that there is significant indication of cluster alignments
in the small separation limit.

\begin{figure}[t]\label{fig:alin}
\psfig{figure=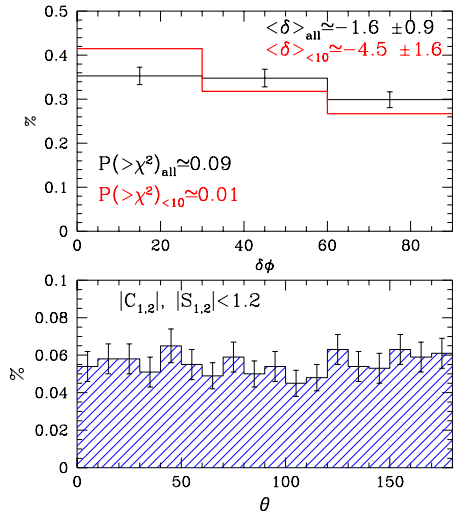,height=2.8in}
\parbox[b]{3.5cm}{
\caption{Upper panel: The distribution of relative position angles between
nearest-neighbours. Lower panel: The distribution of APM cluster
position angles. No orientation bias is present, as seen from the
small values of their Fourier transforms.}}
\end{figure}
In order to test whether this
result is dominated by the ACO cluster pairs, and thus whether it is a
manifestation of the already known Abell cluster alignment effect
(cf. Bingelli \cite{Bin}; Plionis \cite{Plio94}),
we have excluded such pairs to find not only consistent results 
but an even stronger alignment signal. 

Furthermore, we have correlated the alignment signal with the substructure
significance indication in order to see whether there is any relation
between the large-scale environment, in which the cluster distribution
is embedded, and the internal cluster dynamics. 
In figure 4 we present the alignment signal, 
$\langle \delta \rangle$,
between cluster nearest-neighbours (filled dots) and between all pairs
(open dots) with pair separations $< 20$ $h^{-1}$ Mpc. 
There is a strong correlation
between the strength of the alignment signal and the substructure
significance level (see for details Plionis \& Basilakos \cite{Plio02}).

Note that from the analysis of Kolokotronis {\em et al.} \cite{Kol}
it is expected that our procedure will misidentify  
the dynamical state of $\sim 20\%$ of the APM clusters. However, such 
misidentification will act as noise 
and will tend to smear any true alignment-substructure correlation,
since there is no physical reason why random projection effects, within
0.75 $h^{-1}$ Mpc of the cluster core, should be correlated with the
direction of neighbours within distances up to a few tens of
Mpc's (such a correlation could be expected at some level only for
nearest-neighbours in angular space but we have verified that by choosing 
such pairs we obtain an insignificant alignment signal).

\begin{figure}[t]
\psfig{figure=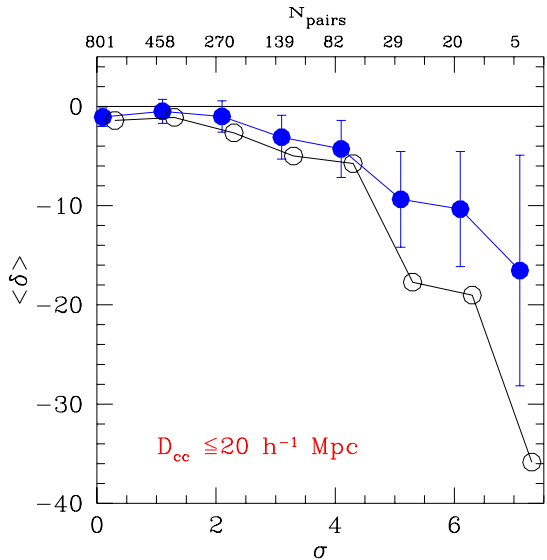,height=2.8in}
\parbox[b]{3.5cm}{\caption{Alignment signal of all 
cluster pairs with separation
$D_{cc}\le 20$ $h^{-1}$ Mpc, as a function of substructure 
significance, $\sigma$. The filled symbols represent the signal
based only on the {\em centroid-shift} substructure categorization
while the open symbols represent the signal from clusters that are also
categorized as having {\em strong} substructure 
by the {\em subgroup} categorization procedure.}}
\end{figure}

\section{Cluster Substructure {\em vs} Local Density}
Our previous results support the formation of clusters by anisotropic 
merging along the filamentary structure within which they are 
embedded (cf. West \cite{We94} \cite{We95}). If
this view is correct then one would expect that clusters with
significant substructure should also reside, preferentially, in
high-density environments (superclusters), and this should then
have an imprint in their spatial two-point correlation function. In the
upper panel of figure 5 we present the spatial 2-point
correlation function of all APM clusters (open dots) and of those with
substructure significance $\sigma \ge 4$ (red dots). It is clear that
the latter are significantly more clustered. This can be seen also in
the insert of figure 5 were we plot the correlation length, $r_{0}$, as a
function of $\sigma$, which is clearly an increasing function of
cluster substructure significance level. To test whether this effect
could be due to the well-known richness dependence of the correlation strength,
we investigated the mean APM richness as a function of $\sigma$ and verified
that if any, there is only a small such richness trend. The conclusion
of this correlation function analysis is that indeed the clusters
showing evidence of dynamical activity reside in high-density
environments, as anticipated from the alignment analysis. It is
interesting that such environmental dependence has also been found 
in a similar study of the BCS and REFLEX clusters 
(Sch\"{u}ecker {\em et al.}, \cite{Sch}) and for
the cooling flow clusters with high mass accretion rates (Loken, Melott
\& Miller \cite{Loken}).
\begin{figure}[t]
\psfig{figure=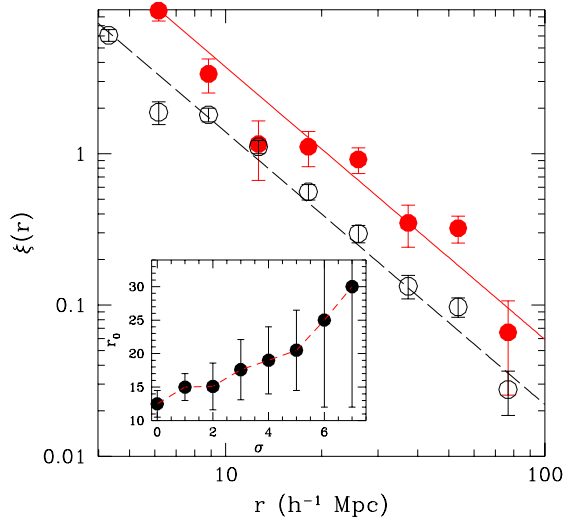,height=2.8in}
\parbox[b]{3.5cm}{\caption{Two-point correlation function of all APM clusters
(open symbols) and of $\sigma>4$ clusters (filled symbols). The lines
represent the best $(r/r_{\circ})^{-1.8}$ fit with $r_{\circ}\simeq12$ and
$\simeq 20$ $h^{-1}$ Mpc respectively. Insert: The cluster correlation length
as a function of substructure significance.}}
\end{figure}

\section{Supercluster - Cluster Alignments}
We have applied a friends of friends algorithm to identify
superclusters in the 3D distribution of APM clusters. We used various
percolation radii and estimated for each supercluster the alignment signal,
$\delta$, estimated between the orientations of all member clusters. 
Furthermore we estimated the mean signal for superclusters of which
all member clusters have substructure
significance $\sigma$ above a chosen threshold.
In figure 6 (left panel) we see that there is a strong correlation
between $\langle \delta \rangle$ and $\sigma$, corroborating the
results of the previous section. In the right panel of figure 6
we present the frequency distribution of the misalignment angle, $\delta\phi$
for all superclusters with more
than one member and the expected Gaussian for a
a uniform distribution. 
We see that indeed there is an excess at small
$\delta\phi$'s. The hatched region shows the corresponding distribution
for those superclusters having all their cluster members with
substructure index $\sigma>2$. It is evident that the distribution is
even more skewed towards small $\delta\phi$'s.
\begin{figure}[t]
\psfig{figure=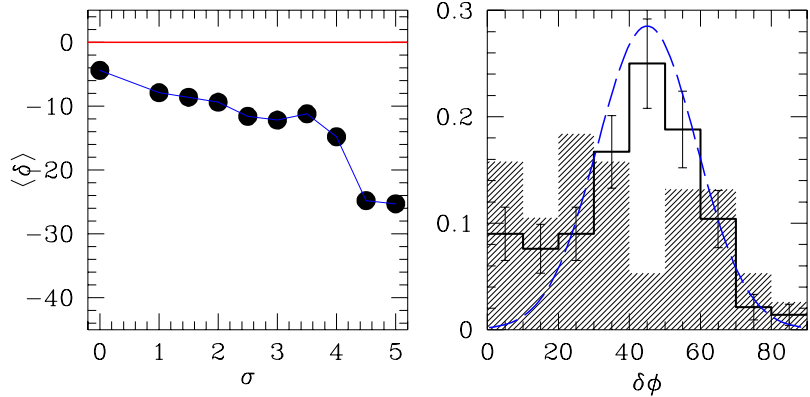,height=2.5in}
\caption{Left panel: Mean alignment signal between supercluster members
versus substructure significance (note that here all members should
have $\sigma$ larger than the indicated limit). Right panel: Frequency
distribution of all supercluster $\delta\phi$ values (histogram), of
those that all member cluster have $\sigma >2$ (hatched region) and the
expected from a uniform distribution (continuous line).}
\end{figure}

We have already established, in section 3, that there is 
an alignment signal between nearby clusters (which is also an
increasing function of cluster substructure significance). A further
interesting question regarding large-scale alignment effects
is whether clusters are also aligned with the orientation 
of their parent supercluster. 
To this end we have estimated the misalignment angle, $\delta\theta$
between the orientation of each supercluster, $\theta_s$, 
with the mean position angle, $\langle \theta \rangle$, 
of its member clusters, ie.;
$\delta\theta = |\theta_s - \langle \theta \rangle|$. In figure 7 we
present the frequency distribution of $\delta\theta$ for two different
supercluster catalogues (based on percolation radii of 20 and 30
$h^{-1}$ Mpc, respectively). The significant excess of small
$\delta\theta$'s is evidence that indeed clusters do
show significant alignments with the orientation of their parent
superclusters. 
\begin{figure}[t]
\psfig{figure=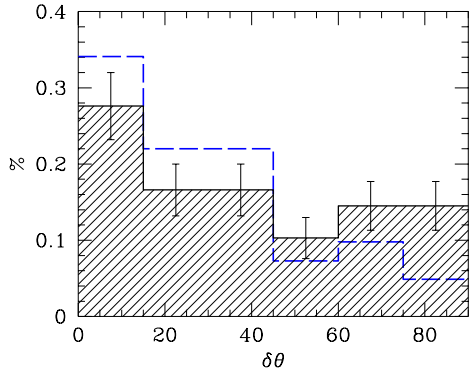,height=2.2in}
\hfill \parbox[b]{4.5cm}{
\caption{Frequency distribution of the misalignment angle between
cluster members and their parent supercluster orientations.
Broken line corresponds to superclusters with percolation radius of
30 $h^{-1}$ Mpc while the hatched distribution to that with percolation
radius of 20 $h^{-1}$ Mpc.}}
\end{figure}

As an individual illustration we present, in figure 8, a filamentary APM
supercluster together with the smooth galaxy density distribution of some
member clusters and the frequency distribution of all the member cluster
position angles. It is evident that there is an excess of clusters with
position angle orientation similar to that of the supercluster itself
(note that filled dots represent clusters with significant substructure).
For the A3112 cluster we overlay also the smooth ROSAT X-ray contours.

\begin{figure}[t]
\begin{center}
\psfig{figure=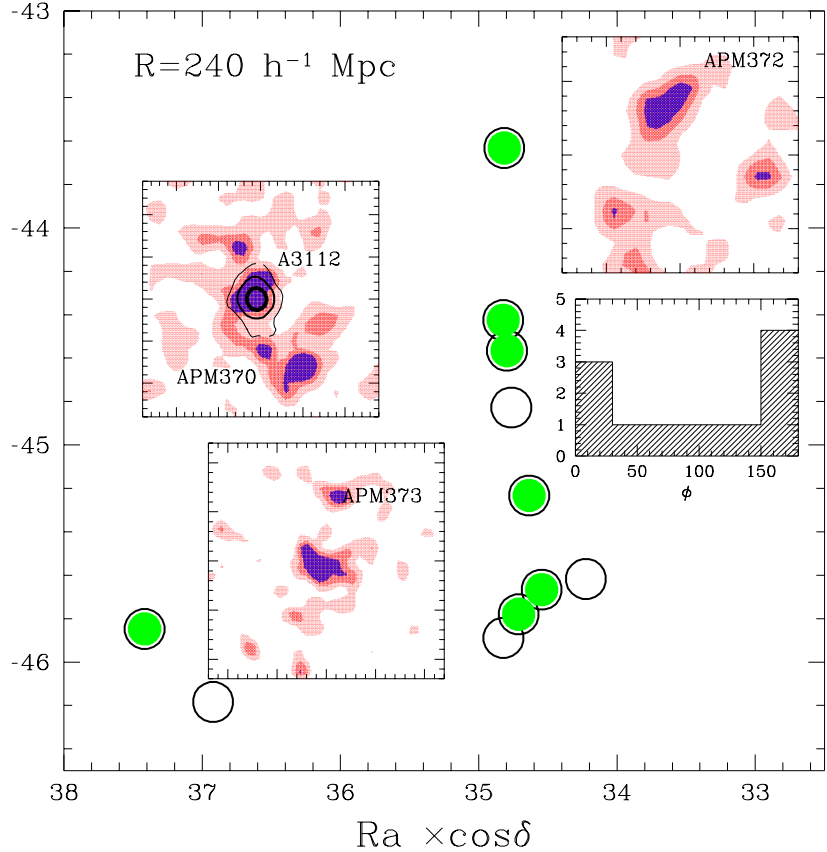,height=4.5in}
\caption{A filamentary APM supercluster containing A3112, A3104,
A3111 as well as 9 poorer APM clusters. The percolation (linking) 
parameter is 12 $h^{-1}$ Mpc. Filled dots represent 
clusters with substructure index $\sigma >2.5$.
There is a
clear tendency of the cluster position angles to be preferentially
aligned with the projected orientation of the supercluster.}
\end{center}
\end{figure}

\section{Conclusions}
We have presented evidence, based on the large APM cluster sample, 
that there is a strong link between the 
dynamical state of clusters and their large-scale environment.
Cluster near-neighbours are statistically aligned with each other and
with the orientation of their parent supercluster. Furthermore,
dynamically young clusters are
significantly more aligned with their nearest
neighbours and they are also much more spatially clustered.
This supports the hierarchical clustering models in which clusters
form by merging along the large-scale filamentary structures within
which they are embedded.

\section*{Acknowledgments}
I thank all my collaborators S.Basilakos, S.Maurogordato and C.Benoist 
for allowing me to present our results prior to publication.


\begin{chapthebibliography}{1}
\bibitem{Arn} Arnaud, M., Maurogordato, S., Slezak, E., Rho, J., 2000,
A\&A, 355, 848
\bibitem{bas00}Basilakos S., Plionis M., Maddox S. J., 2000, MNRAS, 315, 779
\bibitem{Bin}Bingelli B., 1982, AA, 250, 432
\bibitem{Bond86} Bond, J.R., 1986, in {\em Galaxy Distances and
Deviations from the Hubble Flow}, eds. Madore, B.F., Tully, R.B.,
(Dordrecht: Reidel), p.255
\bibitem{Bond87} Bond, J.R., 1987, in {\em Nearly Normal Galaxies},
ed. Faber, S., (New York: Springer-Verlag), p.388
(Dordrecht: Reidel), p.255
\bibitem{borg} Borgani, S. \& Guzzo, L., 2001, Nature, 409, 39
\bibitem{bori}B${\rm \ddot{o}}$hringer H.,
1995, in Proceedings of the 17th Texas Symposium
on Relativistic Astrophysics and Cosmology, eds.
B${\rm \ddot{o}}$hringer H., Tr${\rm\ddot{u}}$mper J., Morfill G. E., 
The New York Academy of Sciences 
\bibitem{Dj}Djorgovski, S., 1983, ApJ, 274, L7
\bibitem{Dal97}Dalton G. B., Maddox S. J., Sutherland W. J., Efstathiou G.,
1997,  MNRAS, 289, 263 
\bibitem{Durr}Durret, F., Forman, W., Gerbal, D., Jones, C., Vikhlinin,
A., 1998, A\&A, 335, 41
\bibitem{Ev}Evrard A.E., Mohr J.J., Fabricant D.G., Geller M.J.,1993, ApJ,
419, L9
\bibitem{Fu}Fuller, T.M., West, M.J. \& Bridges, T.J., 1999, ApJ, 519, 22 
\bibitem{Haa}van Haarlem, M., van de Weygaert, R., 1993, ApJ, 418, 544
\bibitem{Kamp}Kampen van E., Rhee, G.F.R.N., 1990, A\&A, 237, 283
\bibitem{Kol} Kolokotronis, V., Basilakos, S., Plionis, M.,
Georgantopoulos, I., 2001, MNRAS, 320, 49
\bibitem{Lac} Lacey, C., Cole, S., 1996, MNRAS, 262, 627
\bibitem{Loken} Loken, C., Melott, A.L., Miller, C.J., 1999, ApJ, 520, L5
\bibitem{Madd} Maddox S.J. , Sutherland W.J., Efstathiou G., Loveday, J.
1990, MNRAS, 243, 692
\bibitem{Mohr} Mohr, J.J., Evrard, A.E., Fabricant, D.G.,
Geller, M.J., 1995, ApJ, 447, 8
\bibitem{Nov}Novikov, D. {\em et al.}, 1999, MNRAS, 304, L5
\bibitem{Onu} Onuora, L.I., Thomas, P.A, 2000, MNRAS, 319, 614
\bibitem{Plio94}Plionis M., 1994, ApJS., 95, 401
\bibitem{Plio01}Plionis M., 2001, in the proceedings of the
{\em Clusters and the High-Redshift Universe observed in X-rays},
XXI$^{\rm th}$ Moriond Astrophysics Meeting, eds. Neumann et al., 
{\em in press}
\bibitem{Plio02}Plionis M., Basilakos, S., 2001, MNRAS, {\em submitted}
\bibitem{Ri}Richstone, D., Loeb, A., Turner, E.L., 1992, ApJ, 393, 477
\bibitem{Roe93}Roettiger, K., Burns, J. \& Loken, C., 1993, ApJ, 407,
L53
\bibitem{Roe99}Roettiger, K., Stone, J.M., Burns, J., 1999, ApJ, 518, 594
\bibitem{Sar88} Sarazin, C.L., 1988, in {\em X-ray Emission from Clusters
of Galaxies}, Cambridge Astrophysics Series, Cambridge Univ. Press.
\bibitem{Sar01} Sarazin, C.L., 2001, in {\em Merging Processes in
clusters of Galaxies}, eds. Feretti, L., Gioia, M., Giovannini, G., 
(Dordrecht: Kluwer).
\bibitem{Sch99} Schindler S., 1999, in Giovanelli F., Sabau-Graziati
L. (eds.), 
proceedings of the Vulcano Workshop 1999, {\em Multifrequency Behaviour
of High Energy Cosmic Sources}, astro-ph/9909042.
\bibitem{Sch00} Schindler S., 2000, in Giovanelli F., G. Mannocchi (eds.),
proceedings of the Vulcano Workshop 2000, {\em Frontier Objects in 
Astrophysics and Particle Physics}, astro-ph/0010319.
\bibitem{Sch}  (Sch\"{u}ecker, P., Boehringer, H., Reiprich, T.H., 
Feretti, L., A\&A, {\em in press}, (astro-ph/0109030)
\bibitem{Spli}Splinter, R.J., Melott, A.L., Linn, A.M., Buck, C.,
 Tinker, J., 1997, ApJ, 479, 632
\bibitem{Str90}Struble, M.F., 1990, AJ, 99, 743
\bibitem{Str95}Struble, M.F., Peebles, P.J.E., 1985, AJ, 90, 582
\bibitem{Tre}Trevese, D., Cirimele, G., Flin, P., 1992, AJ, 104, 935
\bibitem{We89}West, M. J., 1989, ApJ, 347, 610
\bibitem{We91}West, M. J., Villumsen, J.V., Dekel, A., 1991, ApJ, 369, 287
\bibitem{We94}West, M. J., 1994, MNRAS, 268, 79
\bibitem{We95}West, M. J., Jones C., Forman W., 1995, ApJ, 451, L5
\bibitem{Zab}Zabludoff, A.I. \& Zaritsky, D., 1995, ApJ, 447, L21
\end{chapthebibliography}
\end{document}